\documentclass[11pt]{article}
\usepackage[utf8]{inputenc}
\usepackage{graphicx}
\usepackage[bb=boondox]{mathalfa} 
\usepackage{amsmath, bm} 
\usepackage[dvipsnames]{xcolor}
\usepackage{hyperref} 
\usepackage[toc]{appendix}
\hypersetup{
    colorlinks=true,
    linkcolor=violet,
    filecolor=violet,
    citecolor=violet,
    urlcolor=violet,
}

\usepackage[a4paper, margin=3.1cm]{geometry}

\usepackage{lmodern} 

\title{Parallelizing MCMC Sampling via Space Partitioning}
\author{Vasyl Hafych$^1$, Philipp Eller$^2$, Oliver Schulz$^1$, Allen Caldwell$^1$}
\date{%
    \textit{1: Max Planck Institute for Physics, D-80805 Munich, Germany}\\[0ex]%
    \textit{2: Technical University of Munich, D-80333 Munich, Germany}\\[3ex]%
    \today
}

\begin{document}

\maketitle

\begin{abstract}
Efficient sampling of many-dimensional and multimodal density functions is a task of great interest in many research fields. We describe an algorithm that allows parallelizing inherently serial Markov chain Monte Carlo (MCMC) sampling by partitioning the space of the function parameters into multiple subspaces and sampling each of them independently. The samples of the different subspaces are then reweighted by their integral values and stitched back together. This approach allows reducing sampling wall-clock time by parallel operation. It also improves sampling of multimodal target densities and results in less correlated samples. Finally, the approach yields an estimate of the integral of the target density function.
\end{abstract}

\section{Introduction}

Markov chain Monte Carlo (MCMC) is a technique that allows generating samples with a distribution proportional to a given target density function\footnote{In the following, we will use the terms `target density function' also for unnormalized positive definite continious functions.}. This technique is widely used in Bayesian statistics, statistical mechanics, computational biology, and many other fields or research. One of the major strengths of this technique is that it can converge to the target density even if target functions are highly multidimensional and multimodal. A major difficulty is that convergence is reached only asymptotically, and approaching the stationary distribution can require a very large number of sampling steps. 

A Markov chain, by definition, consists of a series of consecutive steps that move a point or a set of points across the parameter space, which is an inherently serial process. A proposed displacement, by definition of the Markov process, does not depend on the history that led to the current location. Determining whether to accept a displacement involves evaluating the target density at the proposed location.  For many real-world applications, evaluation of a target density can be very computationally costly, and there is usually a limit to how far a single target evaluation can be parallelized efficiently; this can make MCMC sampling very costly. A further complication stems from the fact that a large number of burn-in steps (the steps necessary for the MCMC to reach the stationary distribution) need to be performed for each MCMC chain before representative samples can be generated. The burn-in duration can even exceed the sampling time, especially for target densities that have a complex shape. While separate MCMC chains can be run independently and in parallel, simply increasing their number while producing fewer samples from each chain is therefore not an effective parallelization strategy as the length of the burn-in process for each chain would not change.

Significant research has been conducted to enhance the efficiency of MCMC methods. The developments in this field can be divided into several categories~\cite{ref:robert}. The first is based on exploiting the geometry of the target density function. Hamiltonian Monte Carlo~\cite{ref:hmc} (HMC) belongs to this category and it introduces the auxiliary variable, called momentum, which is refreshed by using the gradient of the density function. 
Symplectic integrators of different orders of precision are developed to approximate Hamiltonian equation~\cite{ref:leapfrog, ref:tslf}. 
The HMC provides less correlated samples than the Metropolis-Hastings algorithm; however, the gradient of the density is not always readily available or cannot be computed in reasonable time. 
The second approach of accelerating MCMC is based on improving the proposal function. Techniques such as simulated tempering~\cite{ref:sim_temp, ref:sim_temp_2}, adaptive MCMC~\cite{ref:adapt_sampling}, and multi-proposal MCMC \cite{ref:multi_try, ref:multi_try_2} are available and have been shown to be effective for many applications~\cite{ref:multi_try_3, ref:sim_temp_3, ref:sim_temp_4, ref:sim_temp_5, ref:adapt_sampling_2}. The third approach is based on breaking initially complicated problems into simpler pieces. For example, 
separate MCMC chains explore the parameter space in parallel and the resulting samples are merged together~\cite{ref:dist_sampling_1, ref:dist_sampling_2}. As discussed earlier, this approach does not simplify convergence of chains to the stationary distribution. 

In this paper, we present an approach that improves MCMC sampling by partitioning the parameter space of the target density function into multiple subspaces and separately sampling the different subspaces. An effective partitioning can thereby change the task from sampling from a complicated target distribution to sampling from many, simpler target distributions.  Any MCMC sampling algorithm can be used to generate samples in the subspaces, and all subspaces can be sampled independently and in parallel. This allows for massively parallel and distributed execution on multiple processors of multiple computer systems. After sampling, the integrated density of each subspace is calculated and the samples for each subspace are re-weighted correspondingly.

Our approach results in increased MCMC sampling efficiency, yielding samples with reduced correlations. It also reduces the required MCMC burn-in time significantly since the possibly multiple modes of the full target density will ideally lie in separate subspaces; each chain will then only have to sample a unimodal density.
In combination, these benefits result in a shorter total sampling time for each MCMC chain (including burn-in), which makes this approach of running many chains distributed over many subspaces efficient, whereas running many chains over the full density is not (due to constant burn-in overhead).

We start with a brief review of Bayesian data analysis, as this is the primary application that we aim for.  In section~\ref{sec:algorithm}, the general idea of the algorithm and our implementation of it are described.  In section~\ref{sec:performance}, performance of the algorithm is shown using the mixture of four multivariate normal distributions as a target density function. Finally, section~\ref{sec:conclusions} summarizes developments presented in this paper and discusses further directions.

\section{MCMC for Bayesian Data Analysis}

For the given model $M$, parameters $\lambda$, and the data $\mathcal{D}$, Bayes' theorem is defined as 
\begin{equation}
\label{eq:bayes}
    P(\lambda | \mathcal{D},M) = \frac{ P(\mathcal{D}| \lambda,M) P_0(\lambda|M) }{P(\mathcal{D}|M)},
\end{equation}
where $P(\mathcal{D}| \lambda,M)$ denotes the likelihood that is used to update the prior probability density $P_0(\lambda|M)$ of $\lambda$ to the posterior probability density $P(\lambda | \mathcal{D},M) $. The denominator is usually called `evidence' or `marginal likelihood' and it is given by the Law of Total Probability
\begin{equation}
P(\mathcal{D}|M)=\int P(\mathcal{D}|\lambda,M) P_0(\lambda|M) d\lambda.
\end{equation}

\noindent MCMC methods do not require knowledge of the normalization constant, $P(\mathcal{D}|M)$, to generate samples with the correct distribution. However, we require the normalizing constant in each subspace in order to reproduce the correct target distribution by patching together samples from different subspaces.  Our approach thereby relies on being able to accurately calculate $P_i( \mathcal{D}| M) $, where $i$ labels one of the subspaces.  A variety of techniques that allow estimating the evidence of models exist, an overview summary can be found in~\cite{rev:evidence_rev}. These typically require a re-sampling of the target probability density function once modes have been identified and only a few of the methods can be used in a post processing step based on existing samples.  We use the recently published AHMI algorithm~\cite{ref:Caldwell} to evaluate the integrals in each subspace directly from the samples.  This provides the user with the correctly weighted samples,  and allows for a simple calculation of the evidence for the full target distribution. This is relevant for evaluating, amongst other things, the Bayes factor used in model comparisons.

\section{Sampling Parallelization via Space Partitioning}
\label{sec:algorithm}

\subsection{Overview}

\begin{figure}[!t]
    \centering
    \includegraphics{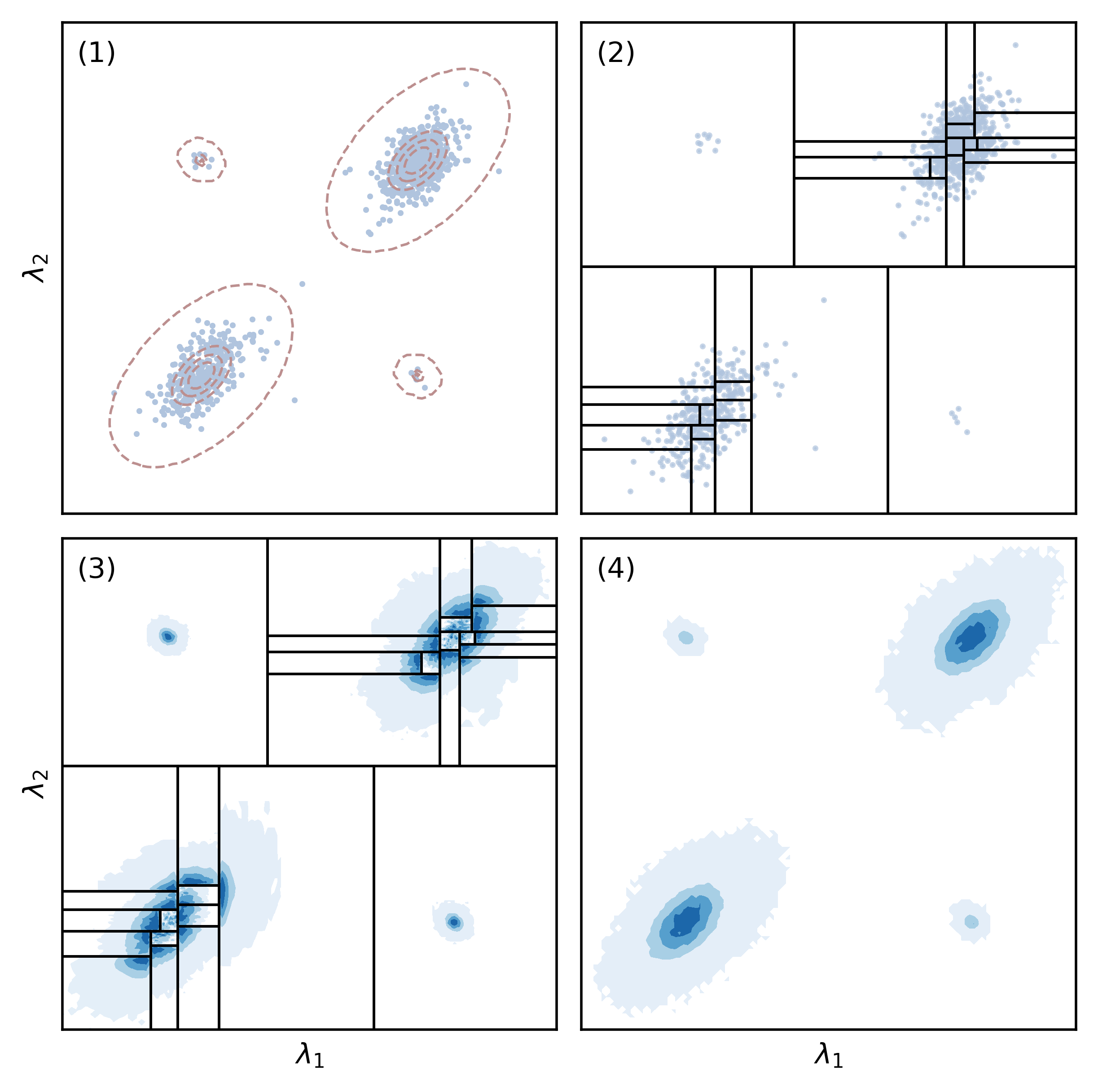}
    \caption{Color maps displaying the steps in the partitioned sampling approach. (1)  The $500$ samples from the exploration step for the target density defined in Eq~\ref{eq:density}. The red dashed lines demonstrate contours of the true density. (2)  The result of partitioning the parameter space into 30 subspaces. The black lines demonstrate the boundaries of subspaces. (3) The $10^4$ samples in each subspace from the individual MCMC chains.  (4) The weighted samples displayed in the full space.}
    \label{fig:example}
\end{figure}

We consider generating samples according to a target density function $f(\boldsymbol{\lambda})$ where $\boldsymbol{\lambda} \subset \mathbb{R}^m$ and $\Omega$ is the support of the function. 
To illustrate our method, we will use as example the sum of four bivariate normal distributions with $\boldsymbol{\lambda} = (\lambda_1, \lambda_2)$:
\begin{equation}
\label{eq:density}
f(\boldsymbol{\lambda}) = \sum_{i=1}^4 a_i \cdot \mathbb{N} (\boldsymbol{\lambda} | \mu_i, \Sigma_i),
\end{equation}
where $a_1 = a_2 = 0.48$, $a_3 = a_4 = 0.02$, $\mu_i = (\pm3.5, \pm3.5)$, $\Sigma_1=\Sigma_2 = (0.33, 0.17; 0.17, 0.33)$, and $\Sigma_3=\Sigma_4 = (0.019, -0.003; -0.003, 0.017)$. Each bivariate normal distribution is individually normalized.  Two, in the upper-right and lower-left quadrants, have large weights ($0.48$) and the other two, in the other quadrants, have small weights ($0.02$).  The covariances are relatively small compared to the separations of the modes, making this a challenging target distribution to sample from for many MCMC algorithms.  Probability contours of this test function are shown in Fig.~\ref{fig:example}-a.

\medskip\noindent Our approach consists of the following four steps: 

\begin{enumerate}
  \item Generate a set of $N_{exp}$ exploration samples $ \left \{ \boldsymbol{\lambda}^{*}_i \right \}_{i=1..N_{exp}} \in \Omega  $, distributed amongst $N_{chains}$, where $N_{exp}$ is a small number compared to the desired number of final MCMC samples and $N_{chains}$ is the number of chains. The chains should have different (possibly randomly chosen) starting points. The samples are used to find region/s of the parameter space with a high density and the MCMC chains are  not required to converge.  An initial sampling of our example function with $N_{exp}=500$ generated using $N_{chains}=25$ with $20$ samples per chain is shown in Fig.~\ref{fig:example}-a.
  
 \item Partition the parameter space into $N_{sp}$ mutually exclusive subspaces $\left \{  \omega_k \right \}_{k=1..N_{sp}} \in \Omega $ in such a way that $\cup \omega_k = \Omega$, $\omega_k \cap \omega_m =\o$ if $ k \neq m $ (see also  Fig.~\ref{fig:example}-b). While in general, the boundaries of the subspaces could be arbitrary shapes, in the following, `$N$ space partitions' will refer to $N$  cuts along different, single parameter axes. This splits the parameter space into $N+1$  rectangular subspaces.
  
  \item Generate $N_{samp}^k$ samples $\left \{  \boldsymbol{\lambda}_i^k \right \}_{i=1..N_{samp}^k}  \in \omega_k$ in each subspace $k$ with the distribution proportional to $f(\boldsymbol{\lambda})$ using a sampling algorithm of choice (see Fig.~\ref{fig:example}-c). 
  Note that each sampler has to perform its burn-in cycle only in the reduced subspace $\omega_k$, which significantly reduces tuning time.  
  
  \item Determine the integrated density of the target distribution in each subspace by computing $ I_k = \int_{\omega_k} f(\boldsymbol{\lambda}) d \boldsymbol{\lambda}  $ and assign the following weights to the sample of subspace $k$ 
  $$w_{k} \propto \frac{I_k}{N_{samp}^k}.$$
  
  \item
  Stitch the now weighted samples together resulting in the final sampling distribution  (see Fig.~\ref{fig:example}-d).

\end{enumerate}

There are many ways of implementing the described idea based on choices of samplers, integrators and space partitioning strategies.
In the following, we describe our implementation that is also made available in the Bayesian Analysis Toolkit package~\cite{ref:github_bat}.  

\subsection{Implementation}
\subsubsection{Exploration samples}

Exploration samples play an important role in this algorithm, since the parameter space is partitioned based on them. If exploration samples represent the structure of the target density closely enough — for example indicating the presence of multiple modes by clusters of spatially neighboring points — then the space partitioning algorithm can capture these features and generate partitions in such a way to split those clusters. This simplifies the target density in each subspace and thus allows for much faster burn-in and tuning procedures. In our implementation, we generate exploration samples by running a large number of MCMC chains, where each chain generates a few hundred samples. There is no tuning or convergence requirement for these chains, but a small set of samples are initially used to set the parameters of the proposal functions for each chain. Some knowledge of the form of the target distribution is useful in determining how many chains and how many samples will be necessary.
While the morphology of the resulting sample clouds should resemble that of the target density as closely as possible, this initial exploration should be fast compared to the following sampling time in the partitioned space.

\subsubsection{Space Partitioning}

\begin{figure}[!t]
    \centering
    \includegraphics{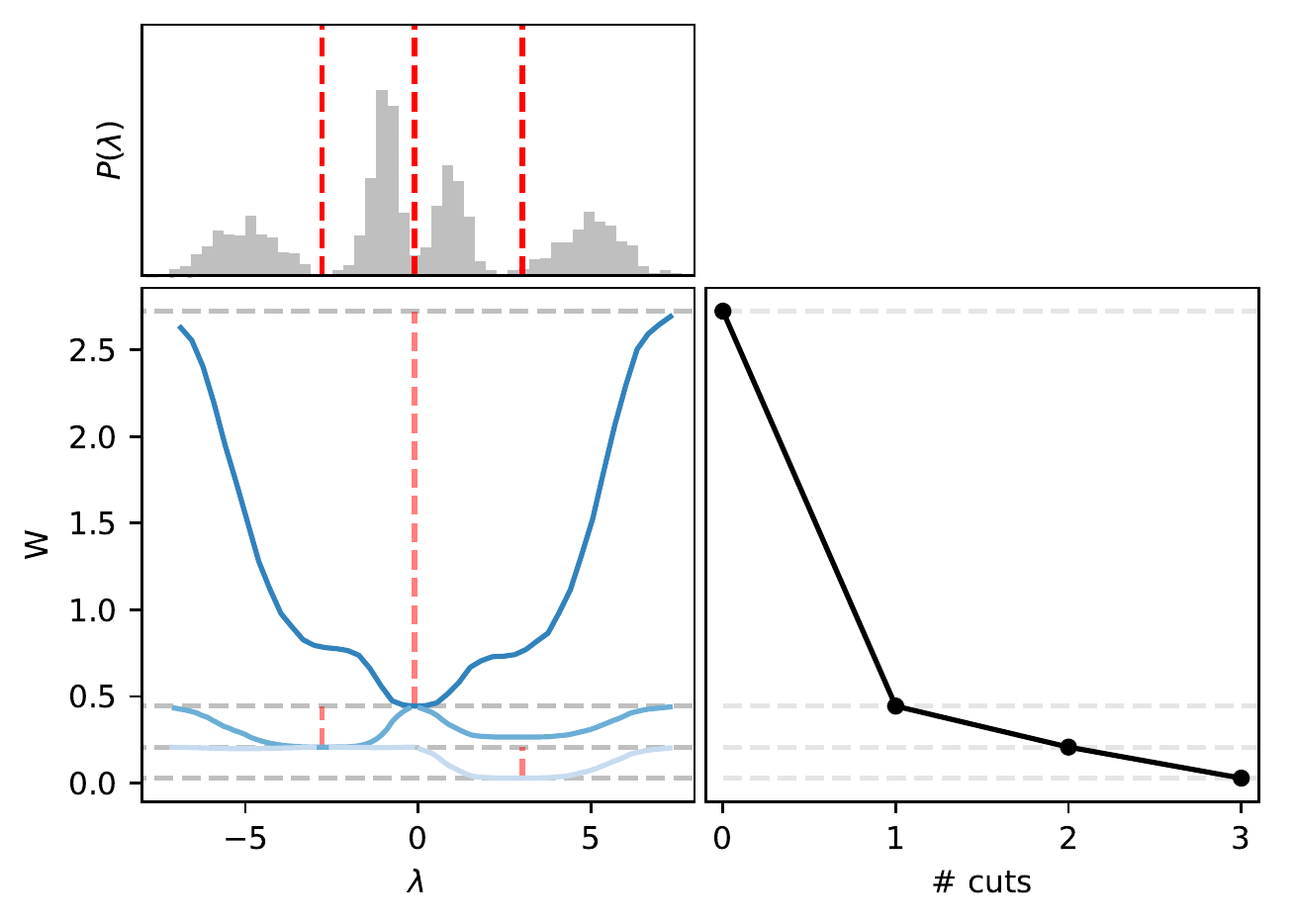}
    \caption{Illustration of the space partitioning algorithm using $5\cdot10^3$ one-dimensional exploration samples $\lambda^{*}$  with a distribution demonstrated in the upper left histogram. The red dashed lines demonstrate the first three cut positions $\widetilde{\lambda}$. The blue lines show the value of $W(a,\lambda)$ as a function of the cut position for 3 iterations of space partitioning. The gray dashed lines show the value of the cost function at its minimum for each iteration. The bottom right subplot illustrates the dependence of $W$ as a function of the number of cuts. }
    \label{fig:space_partitioning}
\end{figure}

Given the discussed exploration samples, we partition our parameter space into rectangular subspaces in such a way as to split clusters of spatially neighboring samples. To do so, a binary tree is used where each node is determined by a cut that is orthogonal to parameter axes.  For the sake of illustration, we consider a one-dimensional problem and exploration samples 
$ \left \{ \lambda \right \}$ (see upper histogram in Fig.~\ref{fig:space_partitioning}). 
The cut position perpendicular to the $\lambda$ axis is denoted as $\widetilde{\lambda}$ and it is selected by finding the minimum of the following cost function:
\begin{equation}
\label{eq:cost_function}
\widetilde{\lambda} 
= \inf_{a}
\left [ 
W(a,\lambda)
\right ] 
=
\inf_{a} 
\left [ 
\sum_{\lambda_i < a} \left | \lambda_i - \left \langle \lambda \right \rangle_{\lambda < a} \right |^2 +
\sum_{\lambda_i > a} \left | \lambda_i - \left \langle \lambda \right \rangle_{\lambda > a} \right |^2 
\right ],
\end{equation}
where $\left \langle \lambda \right \rangle$ denote the mean of samples. This process is then repeated iteratively resulting in the desired number of partitions.

The blue lines in Fig.~\ref{fig:space_partitioning} demonstrate how this cost function depends on the cut positions for 3 partitioning steps.  The partitioning procedure is ended when a minimal change in the cost function results from further partitioning or a maximum number of subspaces is reached. The evolution of the cost function for our example is also shown in Fig.~\ref{fig:space_partitioning}. 

The partitioning procedure is analogous for higher dimensional space. 
If, for instance, a sample vector is $M$-dimensional, then we evaluate Eq.~\ref{eq:cost_function} for every dimension which results in proposed cut positions $\widetilde{\lambda}_i$ with corresponding cost values $W_i(a_i,\widetilde{\lambda}_i)$ for dimension $i=(1,...,M)$.  The minimum cost value is selected and the cut along the corresponding dimension is accepted. 
Additionally, if preliminary knowledge about the structure of the target density is present, the user can specify manually along which parameters partitioning of the parameter space should be performed.

\subsubsection{Sampling}

Sampling in the subspaces is performed independently and does not require communication between MCMC processes. It can therefore be trivially divided into tasks and executed in parallel on multiple processors using distributed computing. In the following, we define a `worker' as a computing unit (this can either be a node of a cluster, networked machine, or a single machine) that consists of multiple CPU cores used to perform one task; i.e., sample the target density in one subspace. All the cores that belong to one worker are called threads and are used to run multiple MCMC chains in parallel within one subspace. Running multiple chains within one subspace is advantageous for determining convergence of the MCMC process. Our implementation allows running subspaces on multiple remote hosts using Julia's support for compute clusters, and MPI/TCP/IP protocols for communication between workers can be used. By default, the Metropolis-Hastings algorithm is used to generate MCMC samples on each subspace. However, any other sampling algorithm can be used. 

\subsubsection{Reweighting}

Samples that originate from different subspaces have different, and a priori unknown, normalizations with respect to each other. In order to correctly stitch those together, a weight proportional to the integral of the target density within the subspace needs to be applied. Given that samples are drawn from the target function $\left \{  \boldsymbol{\lambda}^k \right \} \sim f(\boldsymbol{\lambda}) $ in each subspace $k$, we compute the following integrals:
\begin{equation}
\label{eq:int}
 I_k  =  \int_{\omega_k}  f(\boldsymbol{\lambda}) d\boldsymbol{\lambda}\; .
\end{equation}
As noted earlier, we use the Adaptive Harmonic Mean Integration (AHMI). Given samples drawn according to a density proportional to the function, the AHMI algorithm estimates the integral of the function and the uncertainty of the estimate acting on samples as a postprocessing step, with no requirements to reevaluate the target density. The AHMI algorithm in its current form has been used to accurately integrate functions in up to 20 dimensions.

\subsubsection{Final sample}

Once the weighting is performed, the samples from multiple subspaces are concatenated and returned to the user. The total integral of the function $f(\boldsymbol{\lambda})$ is then estimated by summing the weights of the subspaces $I = \sum_{k=1..N_{sp}} I_k$. 

Our implementation was used to generate the results shown in Fig.~\ref{fig:example}.

\section{Performance}
\label{sec:performance}

In this section, we evaluate the performance of our algorithm on a more complicated test density function. The function was chosen in such a way to (a) have a known analytic integral, (b) allow generating independent and identically distributed (\textit{i.i.d}) samples, and (c) have multiple modes in many dimensions and thus be challenging for a classical Metropolis-Hastings algorithm. With this aim, we have chosen a mixture of four multivariate normal distributions in 9-dimensional space:
\begin{equation}
\label{eq:density}
f(\boldsymbol{\lambda}) = \sum_{i=1}^4 a_i \mathbb{N} (\boldsymbol{\lambda} | \mu_i, \Sigma_i),
\end{equation}
where all $a_i=1/4$ and $\mu$ and $\Sigma$ are randomly assigned mean vector and a covariance matrix.  The full parameter set used for this performance test are given in the Appendix.  Figure~\ref{fig:density} illustrates one and two dimensional distributions of $10^5$ \textit{i.i.d} samples drawn from this density.  

\begin{figure}[!t]
    \centering
    \includegraphics{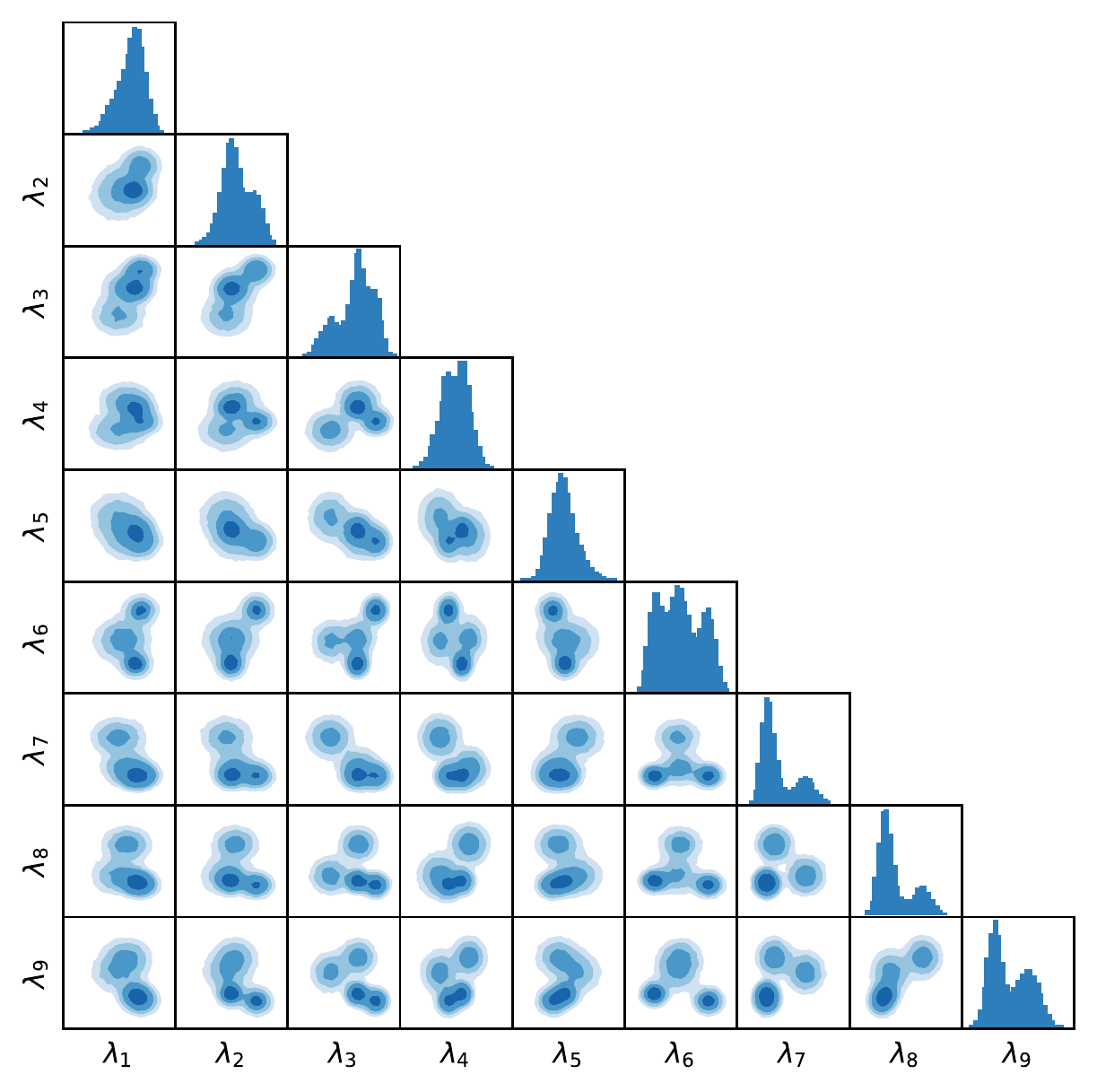}
    \caption{One and two dimensional distributions of the density function given by Eq.~\ref{eq:density}. Histograms are constructed using $10^5$ \textit{i.i.d} samples. }
    \label{fig:density}
\end{figure}

There are two primary points that we demonstrate in this section. The first one is the ability to improve the wall-clock time spent on sampling by utilizing efficiently computational resources. The second one is the ability to improve the quality of samples once we increase the number of space partitions. Measurements of the performance were evaluated as follows: 
\begin{itemize}
    \item  We use a varying number of subspaces $ S=(1, 2, 4, 8, 16, 32)$. Sampling and integration in different subspaces are executed in parallel using 1 worker per subspace with 10 CPU cores per worker. All the CPU cores that are available for the worker are used for multithreaded chain execution. 
    \item In addition, we also vary the wall-clock time that workers can spend on generating samples, considering time intervals of $3, 7, 11$, and $15$ seconds. 
    \item For every combination of space partitions and wall-clock times, we repeat the sampling process 3 times to evaluate statistical fluctuations. 
\end{itemize}
The overall number of MCMC runs is 72.  An example run is: 8 subspaces and 8 workers with 10 CPU cores each are used to sample 10 chains for 11 seconds of wall-clock time, after which samples are integrated and returned; sampling with this setting is repeated 3 times. 

\subsection{Sampling Rate}

\begin{figure}[!t]
    \centering
    \includegraphics{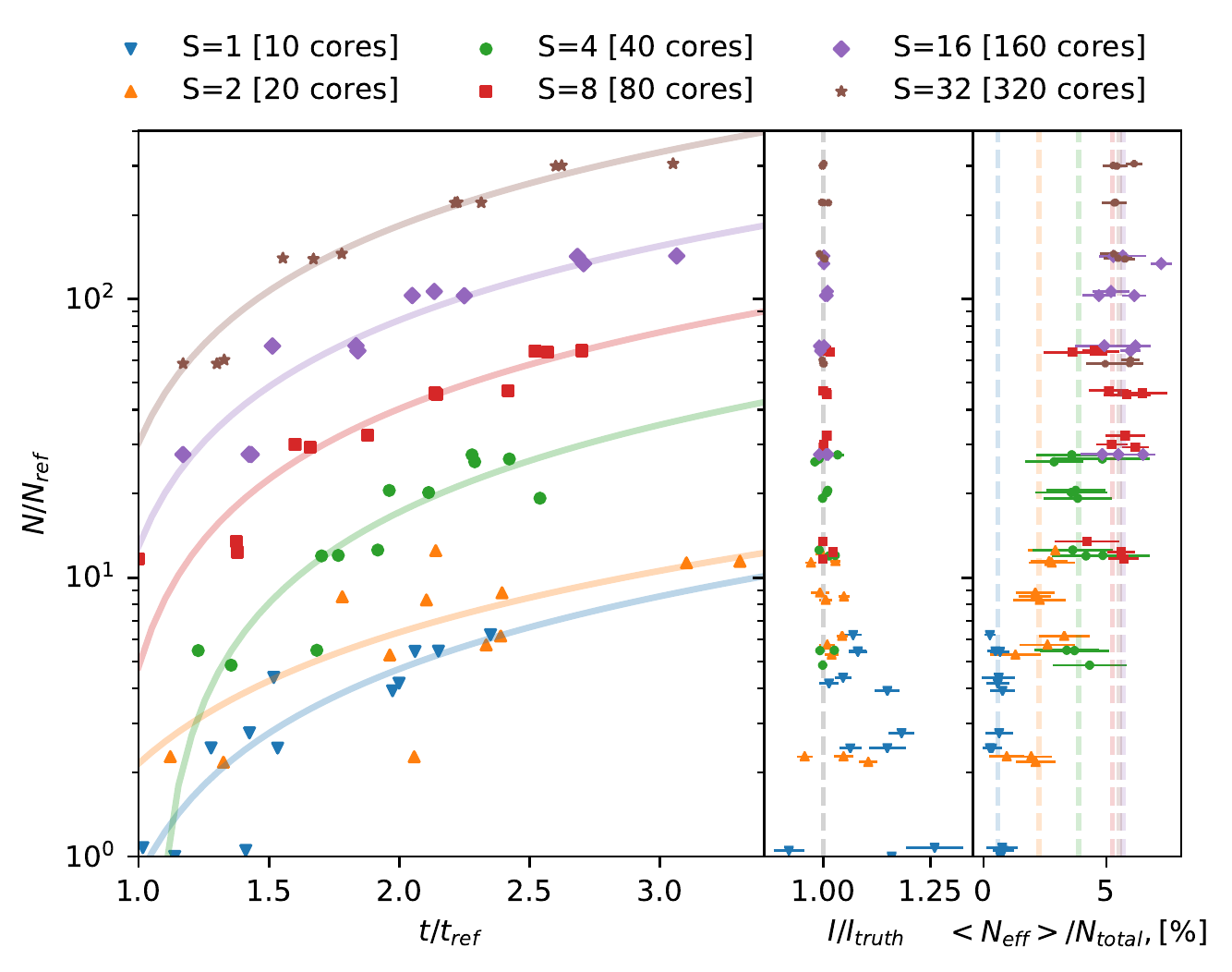}
    \caption{Summary of the benchmark runs for the target density function given by Eq.~\ref{eq:density}. The different colors represent runs with different numbers of partitions; the number of subspaces is denoted by S. The vertical axis is common for all subplots and gives the ratio of the total number of samples generated per single run to the number of samples that are generated if no space partition is performed ($N_{ref} = 3.3\cdot10^4$). \textit{Left subplot}: The horizontal axis shows the ratio of the time spent on sampling and integration to the time that a single worker spent if no space partition is performed ($t_{ref}=14.5$ s). The lines are from linear fits of measurements. \textit{Middle subplot}: The horizontal axis shows the ratio of the integral to the true value; error bars are obtained from the integration algorithm. \textit{Right subplot}: The horizontal axis illustrates the ratio of the effective number of samples (separately for each dimension) to the total number of samples. An effective number of samples is estimated per dimension and the error bars represent the standard deviation across dimensions. Dashed colored lines represent the average fraction over the runs with the same number of space partitions.}
    \label{fig:run_summary}
\end{figure}

A summary of the benchmark run is demonstrated in Fig.~\ref{fig:run_summary}. We used the MPCDF HPC system DRACO\footnote{https://www.mpcdf.mpg.de/services/computing/draco/about-the-system} with Intel `Haswell' Xeon E5-2698 processors (880 nodes with 32 cores @ 2.3 GHz each) to perform parallel MCMC executions. While sampling with space partitions, each subspace requires a different amount of time on sampling and integration, depending on the complexity of the underlying density region. 
The time of the slowest one is reported in Fig.~\ref{fig:run_summary}. 
It can be seen that, by changing the number of subspaces from 1 to 32 (and the number of total CPU cores from 10 to 320), the number of generated samples increases almost two orders of magnitude while the wall-clock time remains constant. 

\begin{figure}[!t]
    \centering
    \includegraphics{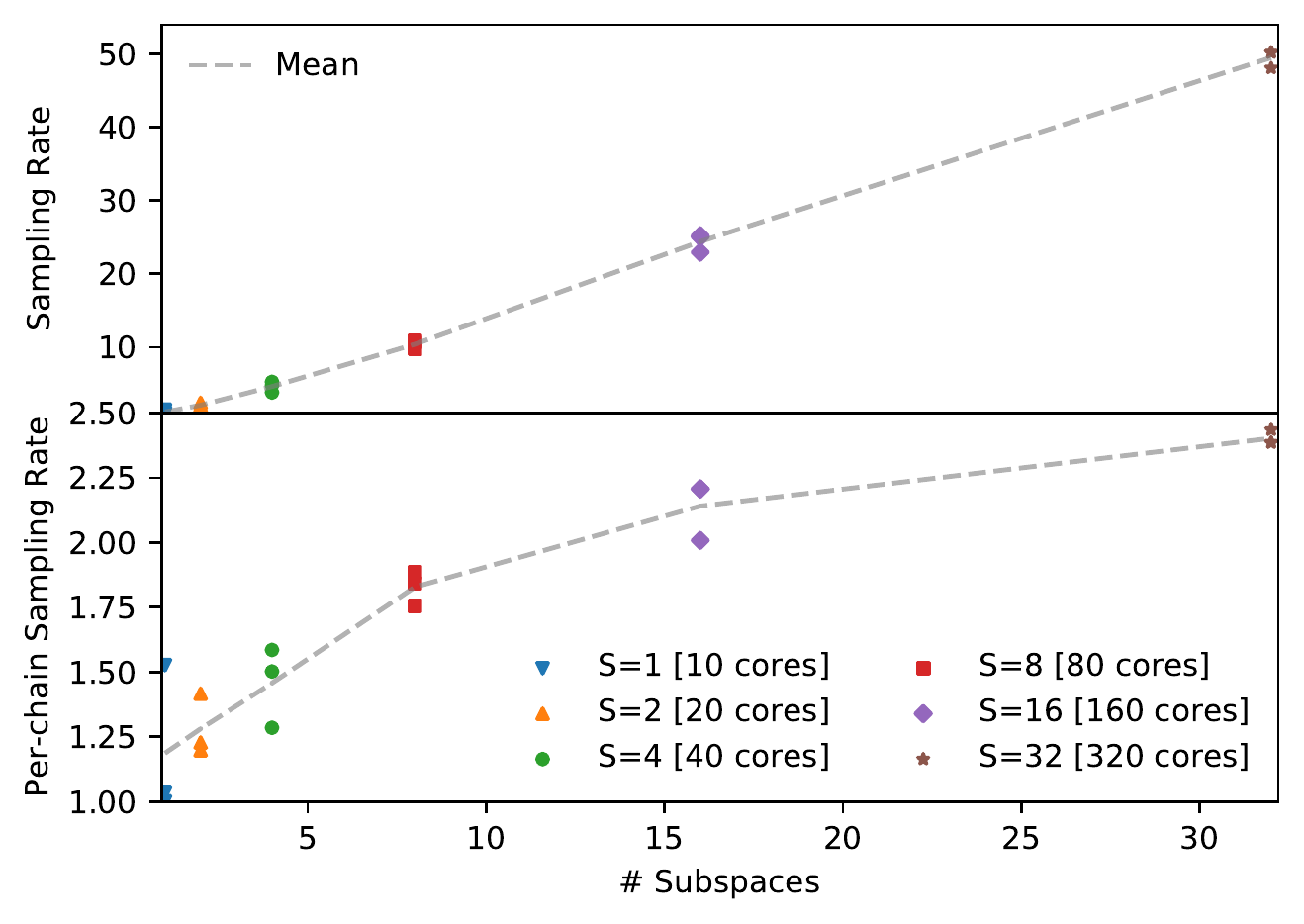}
    \caption{The figure illustrates sampling rate (upper subplot) and per-chain sampling rate (lower subplot) versus the number of space partitions. The gray lines represent average over 3 runs. }
    \label{fig:speedup}
\end{figure}

Figure~\ref{fig:run_summary} can be rearranged into a slightly different form in order to demonstrate the sampling rate.
We define $N_0$ as the number of samples that the sampler with no space partitions has generated  during the time interval $\Delta t_0 = t_{stop} - t_{start}$ ($t_{stop}$ is the wall-clock time when integration has finished and $  t_{start}$ is the wall-clock time when sampling on subspace has started). We further denote as $N_k$ the total number of samples from the run with $k$ subspaces, and the time spent on each subspace as $\Delta t_k$. The sampling rate is defined as
\begin{equation}\label{wc-speedup}
    S = \frac{N_k } {\underset{k}{\textup{max}} \; \Delta t_k} \cdot \frac{ \Delta t_0}{N_0}.
\end{equation}
In addition to the wall-clock time, we measure a CPU time spent on sampling and integration on each subspace using the \textit{CPUTime.jl} package\footnote{A detailed definition of the CPU time can be found in the package documentation.}. We denote it as $\tau_i$, where $i$ is the subspace index, and $\tau_0$ is a CPU time when no space partitions are used. The per-chain sampling rate is defined as
\begin{equation}\label{cpu-speedup}
    S_{per-chain} = \frac{N_k } {\sum_{i=1..k} \tau_i } \cdot \frac{ \tau_0}{N_0}.
\end{equation}
Eq.~\ref{wc-speedup} and Eq.~\ref{cpu-speedup} do not include time spent on the generation of exploration samples and construction of the partition tree. A time spent on the generation of exploration samples depends primarily on the complexity of a likelihood evaluation, and for our problem, it is equal to 4 seconds. The time required to generate the space partition tree primarily depends on the number of exploration samples, it is about 2 seconds for our problem.

The sampling rate and the per-chain sampling rate versus the number of space partitions are presented in Fig.~\ref{fig:speedup}. The figure shows that the sampling rates are improved for both cases. Improvements in the per-chain sampling rate indicates that by partitioning the parameter space we simplify the target density function resulting in faster tuning and convergence (tuning and convergence are occurring in every subspace). While improvement in the sampling rate is expected due to the scaling of the number of CPU cores, its faster-than-linear behavior can be explained by a superposition of the improved per-chain sampling rate and the linear sampling rate. 

\subsection{Density Integration and Effective Sample Size}

Another important characteristic to track is the integral estimate of the target density function. If, for example, samples are not correctly representing the target function, then the integral will deviate from the truth. By partitioning the parameter space we are simplifying tasks for both the sampler and integrator; the complicated problem has been split into a number of simpler ones. This results in better integral estimates, which can be seen in Fig.~\ref{fig:run_summary} (middle subplot).  

Samples that originated from one MCMC chain are correlated. A degree of sample correlation depends on the acceptance probability, number of chains, complexity of the target density function, etc. The effective number of samples can be estimated as 
\begin{equation}\label{ESS}
    N_{eff} = \frac{N}{\hat{\tau}},
\end{equation}
where $N$ is the number of samples and $\hat{\tau}$ is the integrated autocorrelation time, estimated via the normalized autocorrelation function $\hat{\rho}(\tau)$:

\begin{equation}\label{integrated_autocorrelation_time}
    \hat{\tau}_k = 1 + 2 \sum_{\tau = 1}^{\infty} \hat{\rho}_k(\tau)
\end{equation}

\begin{equation}\label{normalized_autocorrelation_function}
    \hat{\rho}_k(\tau) = \frac{\hat{c}_k(\tau)}{\hat{c}_k(0)}
\end{equation}

\begin{equation}\label{c_autocorrelation_function}
    \hat{c}_k(\tau) = \frac{1}{N - \tau} \sum_{n = 1}^{N - \tau} \left( \lambda_{k,i} - \hat{\lambda}_k \right) \left( \lambda_{k,i+\tau} - \hat{\lambda}_k \right)
\end{equation}
where $k$ refers to the one of the M dimensions of the multivariate sample $\boldsymbol{\lambda}_i = \{ \lambda_{1, i}, ...,\lambda_{M, i} \}$ and $\hat{\lambda}_k$ is the average of the $k$-th component of all the $N$ samples. 
In order to compute the sum given by Eq.~\ref{integrated_autocorrelation_time}, we use a heuristic cut-off given by Geyer's initial monotone sequence estimator \cite{ref:Geyer_1992}.
This technique allows us to calculate an effective sample size for each dimension $N_{eff, k} = \frac{N}{\hat{\tau}_k}$. As it is shown in Fig.~\ref{fig:run_summary} (right subplot), the effective number of samples increases with the number of space partitions.  It can also be seen that in our example there is no increase of the fraction of effective samples when the number of subspaces exceeds $8$. 

\subsection{Summary of Scaling Performance}

\begin{figure}[!t]
    \centering
    \includegraphics{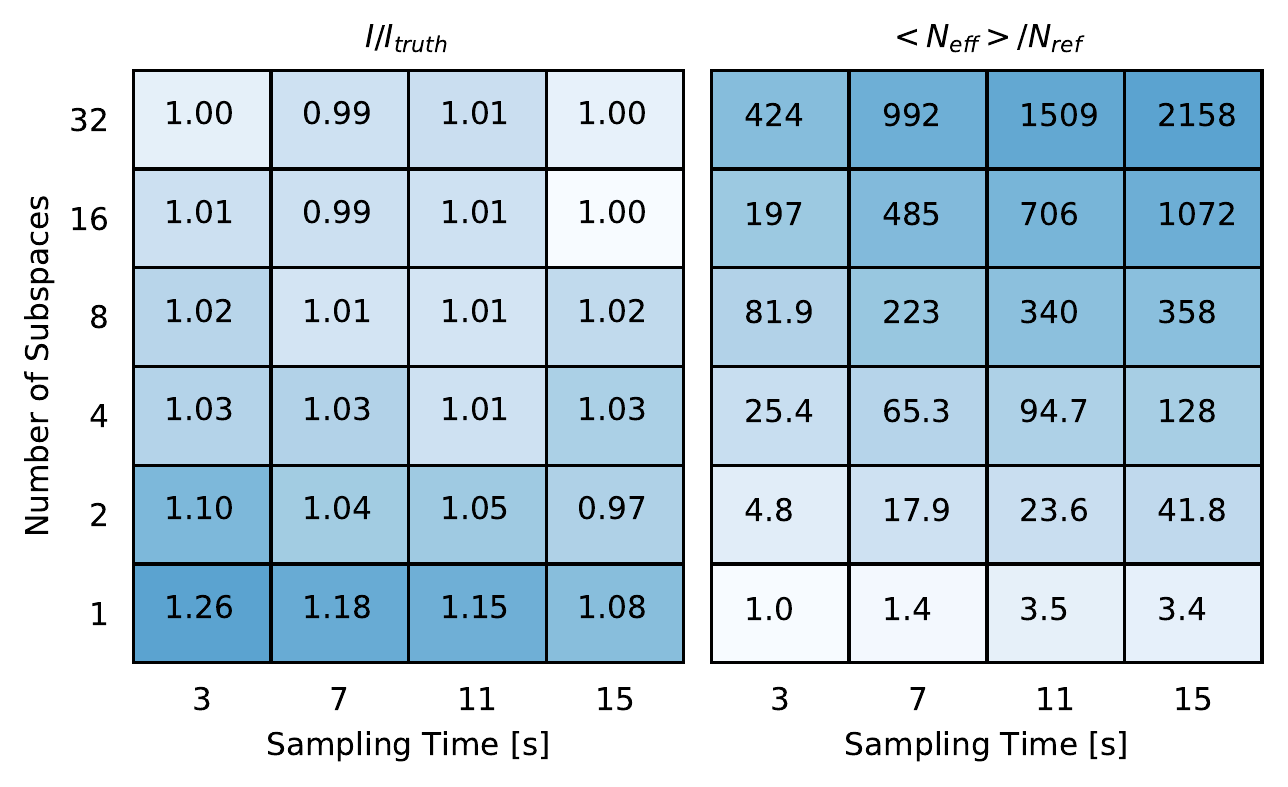}
    \caption{The ratio of the integral to the true value (left panel) and the average number of effective MCMC samples (right panel) for the different numbers of subspaces and sampling times. In the right panel, the values are referenced to the average effective number of samples generated for one subspace and a running time of 3 seconds: $N_{ref} = 267$.}
    \label{fig:table}
\end{figure}

A summary of the performance enhancement from partitioning and sampling in parallel is presented in Fig.~\ref{fig:table}. The accuracy of the integrals of the function for different numbers of subspaces as a function of sampling wall-clock time is shown in the left panel.  There, we see that even with the longest running times tested, the integral estimate without partitioning deviates considerably from the correct value.  Good results are seen already for the shortest running times with 4 subspaces, and running on 32 subspaces gives excellent results in all running times tested.

The right panel in Fig.~\ref{fig:table} shows the average number of effective samples for different combinations of numbers of subspaces and running times.  We find a dramatic increase in the number of effective samples: the factor achieved in the effective number of samples is an order of magnitude larger than the increase in the computing resources (number of processors).  This is due to the much simpler forms of the distributions sampled in the subspaces.

Both of these results show that a strong scaling of the performance is achieved using the partitioning scheme that we have outlined.

\subsection{Sampling Accuracy}

Since our algorithm requires the weighting of samples from different subspaces and stitching them together, we test whether this results in a smooth posterior approximating the true target density function. For comparison, we also approximate the true density by directly generating \textit{i.i.d} samples.
In the following, we will use the two-sample Kolmogorov-Smirnov test~\cite{ref:KS_test} and a two-sample classifier test~\cite{ref:classifier} as a quantitative assessment of how close our MCMC samples are to the \textit{i.i.d} ones. 

\subsubsection{Kolmogorov-Smirnov Test}
\begin{figure}[!t]
    \centering
    \includegraphics{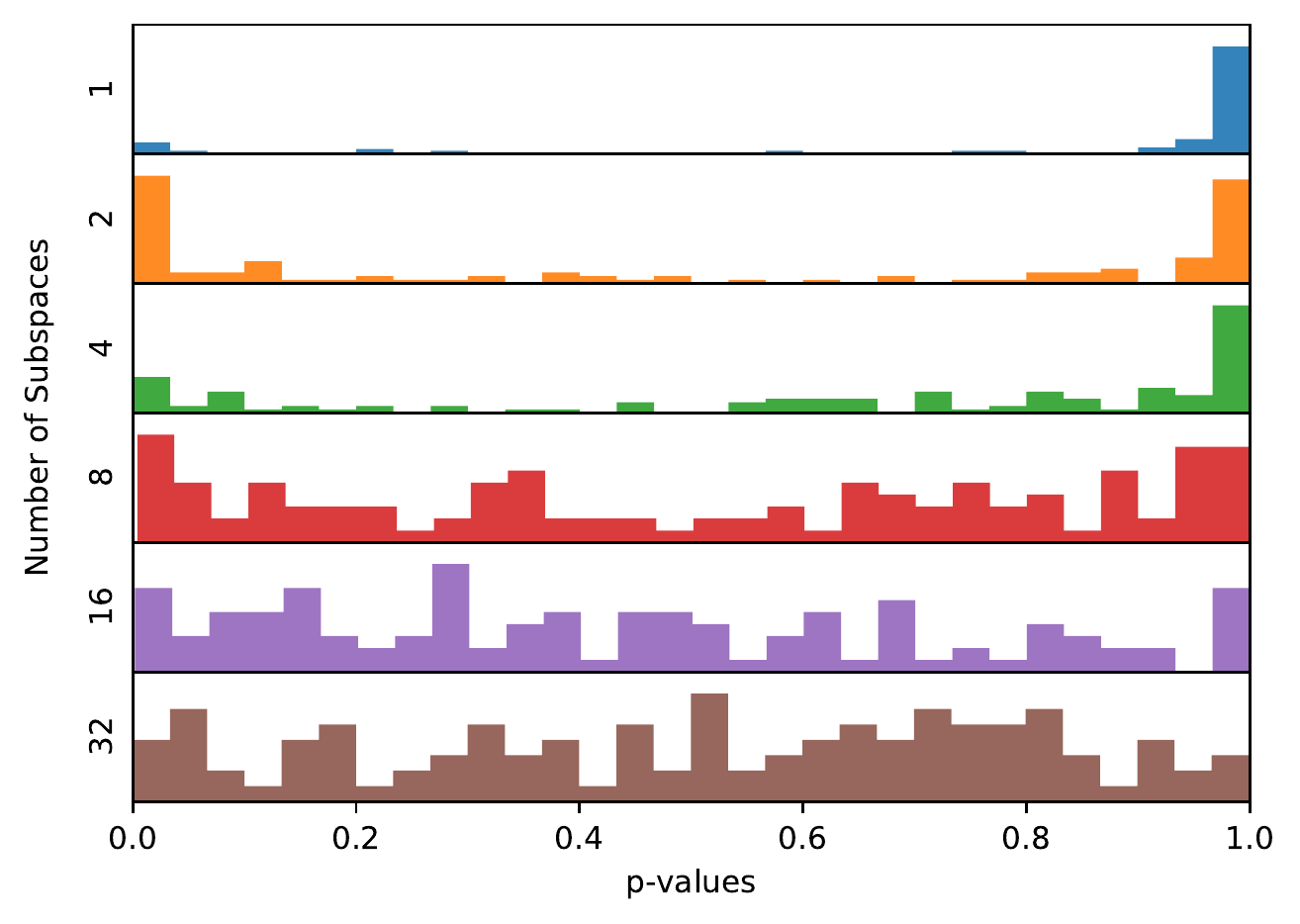}
    \caption{The plot illustrates histograms of the p-values from the two-sample Kolmogorov-Smirnov test to verify whether MCMC and \textit{i.i.d} samples come from the same distribution. Different colors represent different numbers of space partitions. One set of MCMC samples results in 9 p-values for every dimension.}
    \label{fig:ks-test}
\end{figure}

The two-sample Kolmogorov-Smirnov test is used to test whether two one-dimensional marginalized samples come from the same distribution. P-values from this test for every marginal and different number of space partitions are shown in Fig.~\ref{fig:ks-test}.  We use the effective number of samples to calculate p-values for the Kolmogorov-Smirnov test. If two sets of samples stem from the same distribution, then the Kolmogorov-Smirnov p-values should be uniformly distributed. It can be seen that for a small number of space partitions p-values are peaking around 0 and 1.  Peaks close to 1 indicate that the effective number of MCMC samples is likely underestimated, demonstrating that samples are very correlated.  In contrast to this, the peaks near p-values of zero indicate that marginals of \textit{i.i.d}  and MCMC distributions deviate from each other. When using more than $4$ partitions, p-values are uniformly covering the range from $0$ to $1$, indicating that marginals of \textit{i.i.d} and MCMC samples are in a good agreement. 

\begin{figure}[!t]
    \centering
    \includegraphics{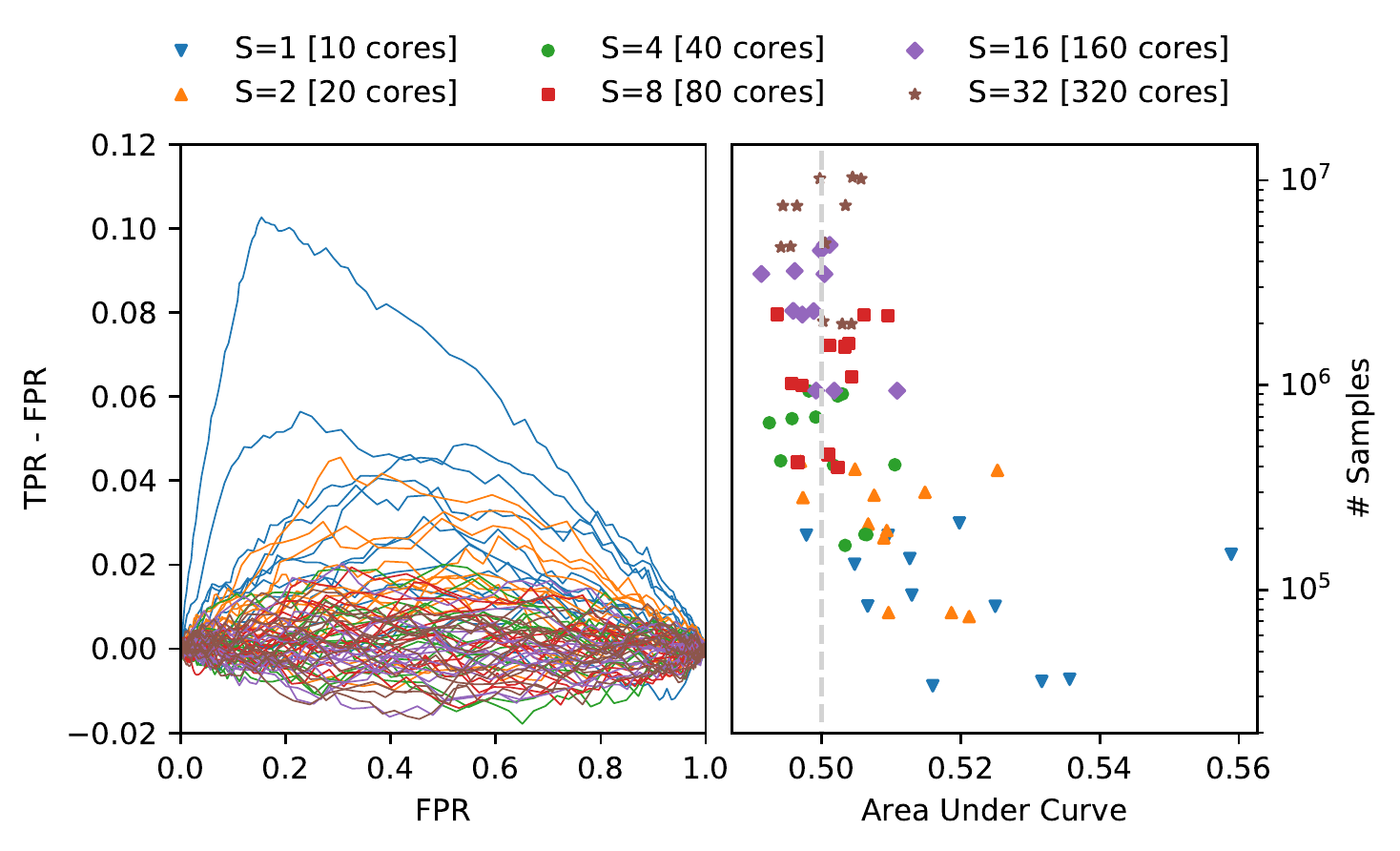}
    \caption{The figure illustrates the results of the classifier two-sample test performed to distinguish \textit{i.i.d} samples and MCMC samples with space partitioning. The left subplot shows a modified receiver operating characteristic (ROC) where TPR stands for true positive rate and FPR for false positive rate. Different lines correspond to a different number of subspaces (S). The right subplot shows area under the ROC curve versus the number of samples.}
    \label{fig:classifier}
\end{figure}

\subsubsection{Classification Test}

A second approach to determining whether two samples are similar to one another uses a binary classifier aiming at distinguishing them. A training dataset is constructed by pairing MCMC and \textit{i.i.d} samples with opposite labels. If samples are indistinguishable, the classification accuracy on the test dataset should be close to that obtained from a random guess. To train a classifier, we use a simple neural network model with two dense layers with sizes $9\times20$ and $20\times2$, and the sigmoid activation function.
To construct training and testing datasets, we generate MCMC and \textit{i.i.d} samples with equal weights. By default, \textit{i.i.d} samples come with weight one. For our weighted MCMC however, we generate $3\cdot10^4$ samples with unit weights by using ordered resampling implemented in~\cite{ref:github_bat}. In total, $4\cdot10^4$ samples are used for training and $2\cdot10^4$ for testing with an equal fraction of MCMC and \textit{i.i.d} samples. Training is performed for every MCMC run that is described in Fig.~\ref{fig:run_summary} individually and results are presented in Fig.~\ref{fig:classifier}. The left subplot in Fig.~\ref{fig:classifier} shows the modified receiver operating characteristic (ROC), where the vertical axis is a difference between true positive rate (TPR) and false positive rate (FPR) for different MCMC runs. If the classifier cannot distinguish two samples, then the line will fluctuate around zero. The right subplot in Fig.~\ref{fig:classifier} shows the integral under the ROC curve (expected to be close to 0.5 for indistinguishable distributions) versus the number of MCMC samples (before resampling was performed). It can be seen that even though the training datasets consist of the same number of MCMC samples, there is a difference in their distinguishability. Samples obtained from the runs with a large number of space partitions have ROC curve integrals much closer to 0.5. It was not possible to detect this difference in the 1d Kolmogorov-Smirnov tests.  

\section{Conclusions}
\label{sec:conclusions}

We have presented an approach to both improve and accelerate MCMC sampling by partitioning the parameter space of the target density function into multiple subspaces and sampling independently in each subspace. These subspaces can be sampled in parallel and the resulting samples then stitched together with appropriate weighting.  The scheme relies on a good space partitioning, which we achieve using a binary partitioning algorithm, and a good integrator for determining the weights assigned to the samples in the different subspaces.  The integrations in our examples were performed using the AHMI~\cite{ref:Caldwell} algorithm.
The parallized MCMC sampling algorithm described in the paper has been implemented in the Bayesian Analysis Toolkit~\cite{ref:github_bat}.

We have benchmarked this technique by evaluating the quality of samples and the sampling rate for a mixture of four multivariate normal distributions in 9-dimensional space. We demonstrate that the space partitioning allows us to obtain a 50-fold increase in the sampling rate while increasing the number of CPUs by a factor 32. This increase is a superposition of two effects: a linear scaling with the number of CPU cores, and a CPU-time reduction due to the simplification of the target density function.  In addition to the increase in the sampling rate, sampling with space partitioning also resulted in an increased quality of MCMC samples by reducing their correlations. This was evidenced in particular by more accurate integral values of the target density.

We have evaluated the correctness of the resulting sampling distributions by comparing the MCMC samples with \textit{i.i.d} samples using a two-sample  Kolmogorov-Smirnov test and with a two-sample classifier test. Both show that increasing the number of space partitions leads to a better agreement between MCMC and \textit{i.i.d} samples. 

Finally, this approach provides the user with a normalization constant of the target density function - the Bayesian evidence is provided at no extra cost.

\clearpage 
\appendix
\section{Appendix}
Covariance matrices used in Eq.~\ref{eq:density} are diagonal with the size $9\times9$ and are equal to
\begin{equation}
\begin{split}
    \Sigma_1 = \textup{diag} (12.64,  12.64, ...,  12.64), \\
    \Sigma_2 = \textup{diag} (10.48,  10.48, ...,  10.48), \\ 
    \Sigma_3 = \textup{diag} (33.03,  33.03, ...,  33.03), \\
    \Sigma_4 = \textup{diag} (27.45,  27.45, ...,  27.45). \\
\end{split}
\end{equation}
Mean vectors are
\begin{align}
\begin{split}
    \mu_1 &= (4.6,14.8,12.7,0.4,-7.3,14.5,-14.0,-9.8,-12.3) , \\
    \mu_2 &= (2.5, 2.9, 2.7, 8.7, -1.6, -11.0, -14.0, -7.5, -8.7), \\ 
    \mu_3 &= (-4.8, 0.68, -12.0, -5.0, 4.4, -0.45, 8.7, -4.5, 2.8), \\
    \mu_4 &= (-1.1, 4.8, 3.3, 13.0, -4.6, 0.99, -9.5, 14.0, 11.0). \\
\end{split}
\end{align}


\begin{thebibliography}{99}

\bibitem{ref:robert} Robert, Christian P., et al. "Accelerating MCMC algorithms." Wiley Interdisciplinary Reviews: Computational Statistics 10.5 (2018): e1435.

\bibitem{ref:hmc} Duane, Simon, et al. "Hybrid monte carlo." Physics letters B 195.2 (1987): 216-222.

\bibitem{ref:leapfrog} Leimkuhler, Benedict, and Sebastian Reich. Simulating hamiltonian dynamics. Vol. 14. Cambridge university press, 2004.

\bibitem{ref:tslf} Blanes, Sergio, Fernando Casas, and J. M. Sanz-Serna. "Numerical integrators for the Hybrid Monte Carlo method." SIAM Journal on Scientific Computing 36.4 (2014): A1556-A1580.

\bibitem{ref:sim_temp} Geyer, Charles J. "Markov chain Monte Carlo maximum likelihood." (1991).

\bibitem{ref:sim_temp_2} Marinari, Enzo, and Giorgio Parisi. "Simulated tempering: a new Monte Carlo scheme." EPL (Europhysics Letters) 19.6 (1992): 451.

\bibitem{ref:adapt_sampling} Douc, Randal, et al. "Convergence of adaptive mixtures of importance sampling schemes." The Annals of Statistics 35.1 (2007): 420-448.

\bibitem{ref:multi_try} Liu, Jun S., Faming Liang, and Wing Hung Wong. "The multiple-try method and local optimization in Metropolis sampling." Journal of the American Statistical Association 95.449 (2000): 121-134.

\bibitem{ref:multi_try_2} Bédard, Mylène, Randal Douc, and Eric Moulines. "Scaling analysis of multiple-try MCMC methods." Stochastic Processes and their Applications 122.3 (2012): 758-786.

 \bibitem{ref:multi_try_3} Laloy, Eric, and Jasper A. Vrugt. "High‐dimensional posterior exploration of hydrologic models using multiple‐try DREAM (ZS) and high‐performance computing." Water Resources Research 48.1 (2012).


\bibitem{ref:sim_temp_3} Neal, Radford M. "Sampling from multimodal distributions using tempered transitions." Statistics and computing 6.4 (1996): 353-366.

\bibitem{ref:sim_temp_4} Xie, Yun, Jian Zhou, and Shaoyi Jiang. "Parallel tempering Monte Carlo simulations of lysozyme orientation on charged surfaces." The Journal of chemical physics 132.6 (2010): 02B602.

\bibitem{ref:sim_temp_5} Carter, J. N., and D. A. White. "History matching on the Imperial College fault model using parallel tempering." Computational Geosciences 17.1 (2013): 43-65.

\bibitem{ref:adapt_sampling_2} Nampally, Arun, and C. R. Ramakrishnan. "Adaptive MCMC-based inference in probabilistic logic programs." arXiv preprint arXiv:1403.6036 (2014).



\bibitem{ref:dist_sampling_1} Mykland, Per, Luke Tierney, and Bin Yu. "Regeneration in Markov chain samplers." Journal of the American Statistical Association 90.429 (1995): 233-241.

\bibitem{ref:dist_sampling_2} Jacob, Pierre E., John O'Leary, and Yves F. Atchadé. "Unbiased markov chain monte carlo with couplings." arXiv preprint arXiv:1708.03625 (2017).

\bibitem{ref:mcmc_bayes} Gelfand, Alan E., and Adrian FM Smith. "Sampling-based approaches to calculating marginal densities." Journal of the American statistical association 85.410 (1990): 398-409.

\bibitem{rev:evidence_rev} Friel, Nial and Jason Wyse. “Estimating the evidence – a review.” Statistica Neerlandica 66 (2011): 288-308.

\bibitem{ref:Caldwell} Caldwell, Allen, et al. "Integration with an Adaptive Harmonic Mean Algorithm." arXiv preprint arXiv:1808.08051v2 (2020).

\bibitem{ref:github_bat} The Bayesian Analysis Toolkit repository, \url{https://github.com/bat/BAT.jl}


\bibitem{ref:Geyer_1992} Geyer, Charles J. "Practical markov chain monte carlo." Statistical science (1992): 473-483.

\bibitem{ref:KS_test} Klotz, Jerome. "Asymptotic efficiency of the two sample Kolmogorov-Smirnov test." Journal of the American Statistical Association 62.319 (1967): 932-938.

\bibitem{ref:classifier} Lopez-Paz, David, and Maxime Oquab. "Revisiting classifier two-sample tests." arXiv preprint arXiv:1610.06545 (2016).

\end{thebibliography}
\end{document}